\documentclass[referee,english]{raa}
\usepackage{graphicx,times,epstopdf,color,caption}
\usepackage{natbib}
\usepackage{amssymb,amsmath}
\usepackage{babel}
\usepackage{booktabs}
\usepackage{threeparttable}
\usepackage{hyperref}
\hypersetup{pdftitle = The title of my PDF, pdfauthor = My name, pdfsubject= The subject, pdfkeywords = keyword1 keyword2 keyword3}
\hypersetup{colorlinks = true, linkcolor = green, anchorcolor = red, citecolor = blue, filecolor = red, pagecolor = red, urlcolor = red}

\begin{document}

   \title{Machine Learning Classification of Gaia Data Release 2
}

 \volnopage{ {\bf 20XX} Vol.\ {\bf X} No. {\bf XX}, 000--000}
   \setcounter{page}{1}

   \author{Yu Bai\inst{1}, JiFeng Liu\inst{1,2}, Song Wang\inst{1}
   }
%% Here is an example of three authors come from different institutes.
%% For single author or all the authors from an institute, use "\inst{}" only

   \institute{ Key Laboratory of Optical Astronomy, National Astronomical Observatories, Chinese Academy of Sciences,
       20A Datun Road, Chaoyang Distict, Beijing 100012, China; \it{ybai@nao.cas.cn}\\
%% Please give the E-mail address of the author, to whom future correspondence and
%% offprint requests will be sent.
        \and
             College of Astronomy and Space Sciences, University of Chinese Academy of Sciences, Beijing 100049, China
\vs \no
   {\small Received 20XX Month Day; accepted 20XX Month Day}
}

\abstract{  Machine learning has increasingly gained more popularity with its incredibly
          powerful ability to make predictions or calculated suggestions for large amounts of data.
          We apply the machine learning classification to 85,613,922 objects in the $Gaia$ data release 2,
          based on the combination of the Pan-STARRS 1 and AllWISE data.
          The classification results are cross-matched with Simbad database, and the total accuracy is 91.9\%.
          %With the help of the Simbad database, we set up the criteria for the reliable classifications.
          Our sample is dominated by stars, $\sim$ 98\%, and galaxies makes up 2\%.
          For the objects with negative parallaxes, about 2.5\% are galaxies and QSOs,
          while about 99.9\% are stars if the relative parallax uncertainties are
          smaller than 0.2. Our result implies that using the threshold of 0 $< \sigma_\pi/\pi <$ 0.2
          could yield a very clean stellar sample.
\keywords{methods: data analysis --- stars: general --- Gaia catalog
}
}

   \authorrunning{Y. Bai et al. }            %author_head in even pages
   \titlerunning{Machine Learning Classification of Gaia Data Release 2}  % title_head in odd pages
   \maketitle

%________________________________________________ sections below
%
\section{Introduction}
The ESA space mission $Gaia$ performs an all-sky astrometric, photometric, and radial velocity survey
at optical wavelengths \citep{Gaia16a}. The primary objective of the $Gaia$ mission is to survey more
than one billion stars, in order to investigate the origin and subsequent evolution of our Galaxy.
Its second data release ($Gaia$ DR2; \citealt{Gaia18}) includes $\sim$ 1.3 billion objects with valid
parallaxes. These parallaxes are obtained with a complex iterative procedure, involving various
assumptions \citep{Lindegren12}. Such procedure may produce parallaxes for galaxies and QSOs, which
should present no significant parallaxes \citep{Liao18}.

Besides, $Gaia$ uses two fields of view to observe, and this in principle might lead to a global
parallax bias \citep{van05,Butkevich17,Liao17}.
Separating galaxies and QSOs from stars allows us to characterize the parallax bias in the
$Gaia$ catalog, and to provide a clean and accurate stellar sample for further investigation.
Traditionally, the classification of objects involves magnitudes and colors criteria,
but the criteria become too complex to be described with functions in a multidimensional
parameter space. By contrast, this parameter space can be effectively explored with machine-learning (ML)
algorithms, which have helped us to deal with complex problems in modern astrophysics
\citep{Huertas08,Huertas09,Manteiga09,Bai18a,Pashchenko18}.

ML provides us an alternative option to classify billions of objects that cannot
be followed-up spectroscopically. \citet{Bai18a} applied the supervised ML
to the star/galaxy/QSO classification based on the combination of SDSS and LAMOST spectral surveys (the SL classifier).
Actually, the class labels of the training objects are from spectroscopy, and are regarded as true. 
Narrow line QSOs are classified as galaxies by both SDSS and LAMOST pipeline because the 
template of QSO in the pipelines is the theoretical one with broad emission lines.
The classifier built with the random forest algorithm showed best performance on time cost and the inner accuracy.
Several blind tests were also performed on the objects observed by the RAVE, 6dFGS and UVQS.
The accuracies were higher than 99\% for the stars and galaxies,
and higher than 94\% for the QSOs.

In this paper, we apply the SL classifier to the $Gaia$ DR2 to investigate the
potential extragalactic objects. The data and classification are described
in Section \ref{data}. Section \ref{res} gives the result and analysis, and a summary
is given in Section \ref{sum}.

\section{Data and Classification}\label{data}

In order to use the SL classifier, we build a nine-dimensional color space,
$g-r$, $r-i$, $i-J$, $J-H$, $H-K$, $K-W$1, $W$1$-$$W$2, w1mag\_1$-$w1mag\_3, and w2mag\_1$-$w2mag\_3 \citep{Bai18a}.
The optical colors are extracted from the data release 1 of the Panoramic Survey Telescope and
Rapid Response System (Pan-STARRS 1; hereafter PS1) archive data. The PS1 has carried out a set of imaging sky surveys
including the 3$\pi$ Steradian Survey, in which the mean 5$\sigma$ point source limiting sensitivities
are 23.3, 23.2 and 23.1 mag in $g, r, i$ bands \citep{Chambers16}. We cross-matched the $Gaia$ DR2 with
PS1 using $panstarrs1\_best\_neighbour$, the pre-computed PS1 cross-match table provided in the $Gaia$
archive \citep{Marrese17}. The table includes 810,359,898, the most likely matches between PS1 and
$Gaia$ DR2, which were determined with the angular distances, position errors, epoch differences,
and density of sources in PS1.

In order to obtain the infrared colors, we cross-matched the $Gaia$ DR2 with AllWISE catalog using
$allwise\_best\_neighbour$, which includes 300,207,917 matches \citep{Marrese17}. Here we select
the objects with the S/N ratios higher than 2 in the $W$1 and $W$2 bands.
As a result, the cross-matchings yield 85,613,922 objects with the valid nine colors.
We feed the SL classifier with the nine-color matrix, and the classifier returns the types and the possibilities ($P$) for
stars, galaxies and QSOs. The sum of $P$ for three types is 100\%, and the type with the highest $P$ is adopted by
the SL classifier as the output type. Therefore, the $P$ of the adopted type is higher than 33.33\%.

Traditionally, QSOs are separated from other AGNs mainly by their absolute $B$ magnitudes.
The QSOs in training data of the SL classifier are identified with the QSO spectral templates.
The different definitions of QSOs may cause many galaxies in our sample classified as QSOs in literatures.
Therefore, the galaxies and QSOs given by the SL classifier are combined and hereafter called galaxies.
The results includes 83,891,260 stars, and 1,722,662 galaxies.
%The results includes 83,891,260 stars, 1,614,046 galaxies, and 108,616 QSOs.

\section{Result and Analysis}\label{res}

\subsection{Comparison with Simbad}
\begin{figure}
   \centering
    \includegraphics[width=1\textwidth]{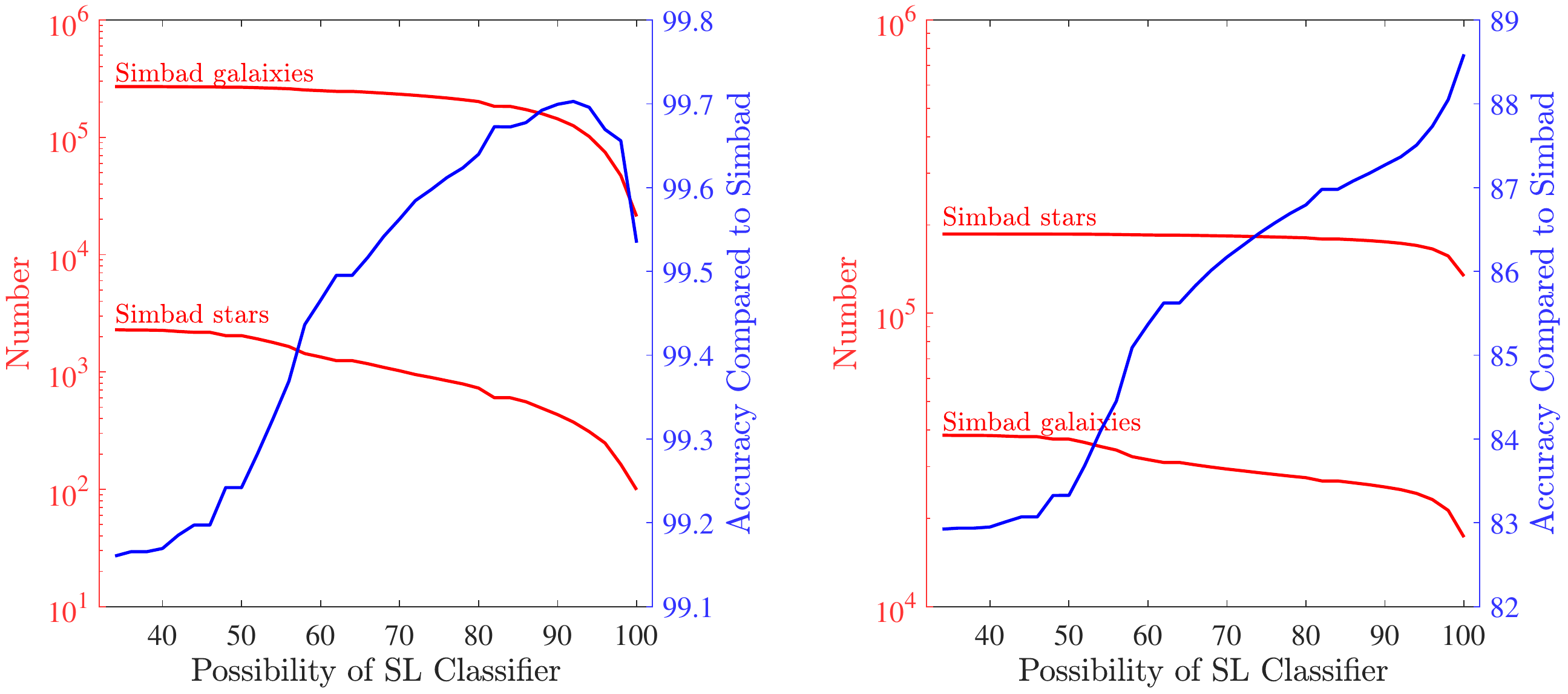}
  \caption{The ML possibility distributions for the galaxies (left panel) and stars (right panel).
           The red lines are the numbers of Simbad classifications higher than the corresponding possibilities.
           The y axis is in log scale for clarity.
           The blue lines are the accuracies compared to Simbad. }
  \label{sco}
\end{figure}

We cross-match these objects to the Simbad database in order to estimate the probability of the possibly wrong classifications.
%The QSOs are separated from other AGNs mainly by their absolute $B$ magnitudes in the Simbad database, while
%the QSOs in training data of the SL classifier are identified with the QSO spectral templates.
%The different definitions of QSOs may cause many galaxies in our sample classified as Simbad QSOs.
%Therefore, the galaxies and QSOs are cross-matched together.
The Simbad database gives 308,864 galaxies, 191,497 stars and 10,987 unclassified objects.

The distributions of the output possibilities are presented in Fig. \ref{sco}.
We defined the accuracies as the ratios between the numbers of the Simbad types and those given by SL
classifier. The total accuracy is 91.9\%.
More than 99.1\% of the galaxies in our sample are also classified as galaxies in the Simbad database,
and more than 83\% of the stars in our sample are classified as stars.

%The accuracy reaches 99.7\% for the galaxies with $P >$ 92\%, and the
%accuracy is 88.6\% for the stars with $P =$ 100\%.
%In order to obtain more reliable classifications,
%we apply the criteria to our sample, $P >$ 92\% for galaxies and QSOs and $P =$ 100\% for the stars,
%which yields 48,079,284 stars, 367,644 galaxies and 34,718 QSOs.

The classification accuracy of the stars is lower than those of the spectrally resolved samples in \citet{Bai18a}.
The stars in the training sample of the SL classifier are mainly from LAMOST, which is dominated by the
stars located in the Galactic anticenter. This selection effect may make the SL classifier familiar with the
lightly reddened objects. The objects located at the heavily
reddened direction of the Galaxy are probably hard to be recognized by the SL classifier.

%The main types of the Simbad database could yield the wrong classification \citep{Lesteven95}.

\subsection{Sky Distribution}\label{sd}
%\begin{figure}
%   \centering
%    \includegraphics[width=1\textwidth]{galgb.pdf}
%  \caption{The density distributions of the classification results in Galactic coordinates. \textbf{Left panel:
%           the objects that satisfy our criteria of $P =$ 100\% for the stars and $P >$ 92\% for the galaxies and QSOs.}
%           Righ panel: the objects that don't satisfy the criteria.
%           The colorbar indicates the number of objects per deg$^2$. }
%  \label{galgb}
%\end{figure}

\begin{figure}
   \centering
    \includegraphics[width=1\textwidth]{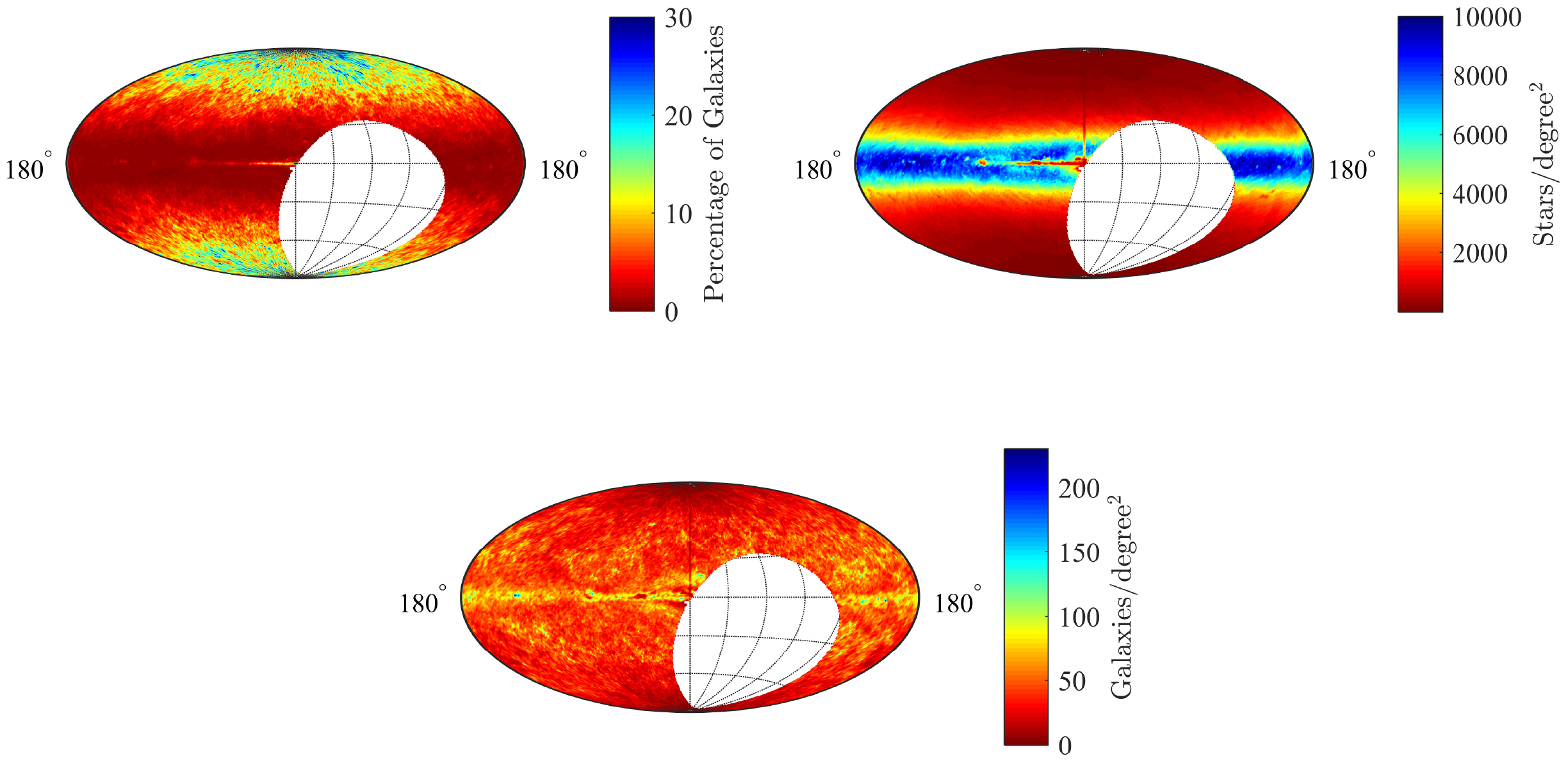}
  \caption{The distributions of the classification results in Galactic coordinates. Upper left panel: the percentage of the galaxies per degree$^2$.
           Upper right panel: the density of the stars. Lower panel: the density of the galaxies. }
  \label{gal}
\end{figure}

%The left panel of Fig \ref{galgb} shows the distribution of the 48,481,646 objects that meet our criteria, and
%the right panel shows the distribution of the 37,132,276 objects that do not meet the criteria.
%The objects in the right panel are probably associated with Galactic clouds, if comparing their distribution to
%the IRAS 100 micron imaging data \citep{Juvela12}. The high extinction can severely redden the optical
%colors. These highly reddened colors are probably located outside the parameter space of the training data for the SL
%classifier, since the limit magnitude of $Gaia$ is 20.7 mag in the $G$ band, fainter than the spectrally resolved objects
%in LAMOST and SDSS \citep{Gaia16b}. Additionally, the $WISE$ photometry is limited by confusion near the Galactic
%plane due to high source density \citep{Wright10}.
%The SL classifier returns relatively low possibilities implying that the nature of these objects are hard to be recognized
%with these optical and infrared colors.

We present distributions of the classification results in Fig. \ref{gal}.
It is expected that the Galactic plane is dominated by the stars, and the percentages of the galaxies become higher at high latitudes.
The relative high percentages in the most central Galactic plane may be due to the low density of
the stars in this region (left panel in Fig. \ref{gal}).
The low completeness of PS1 caused by the high extinction \citep{Chambers16} maybe results in
the low density of the stars in the most central Galactic plane.
%Since the completeness of PS1 is low along the Galactic plane \citep{Chambers16},
%the density of the stars is lower in the most central Galactic plane.
Additionally, the $WISE$ photometry is limited by confusion near the Galactic
plane due to high source density \citep{Wright10}.
%The high extinction of the Galactic clouds could lead to low $P$ of stars,
%and result in the lower density in the most center of the Galactic plane.
In the distribution of the galaxies, we can find over-density areas corresponding to some galaxy clusters \citep{Jarrett04},
e.g., Abell 624, Perseus-Pisces supercluster, and Shapley concentration.

%The average density of the QSOs at high Galactic latitudes ($b > 15\degr$ or $< -15\degr$) is $\sim$ 1 per square degree,
%smaller than the density in \citet{Liao18} and \citet{Paine18}. Their catalogs include other types of AGNs rather than QSOs,
%while these AGNs may be classified as galaxies in our result. On the other hand, the $Gaia$ and PS1 photometry is deeper
%than the 2MASS and $WISE$ photometry, and the source selection based on AllWISE data is probably inefficient
%to detect the deep sky QSOs. That is why the density of $Gaia$-WISE AGNs \citep{Paine18} is extremely lower than that of
%multi-bands selected AGNs \citep{Liao18}.

\subsection{Relative Error of Parallax}
\begin{figure}
   \centering
    \includegraphics[width=.9\textwidth]{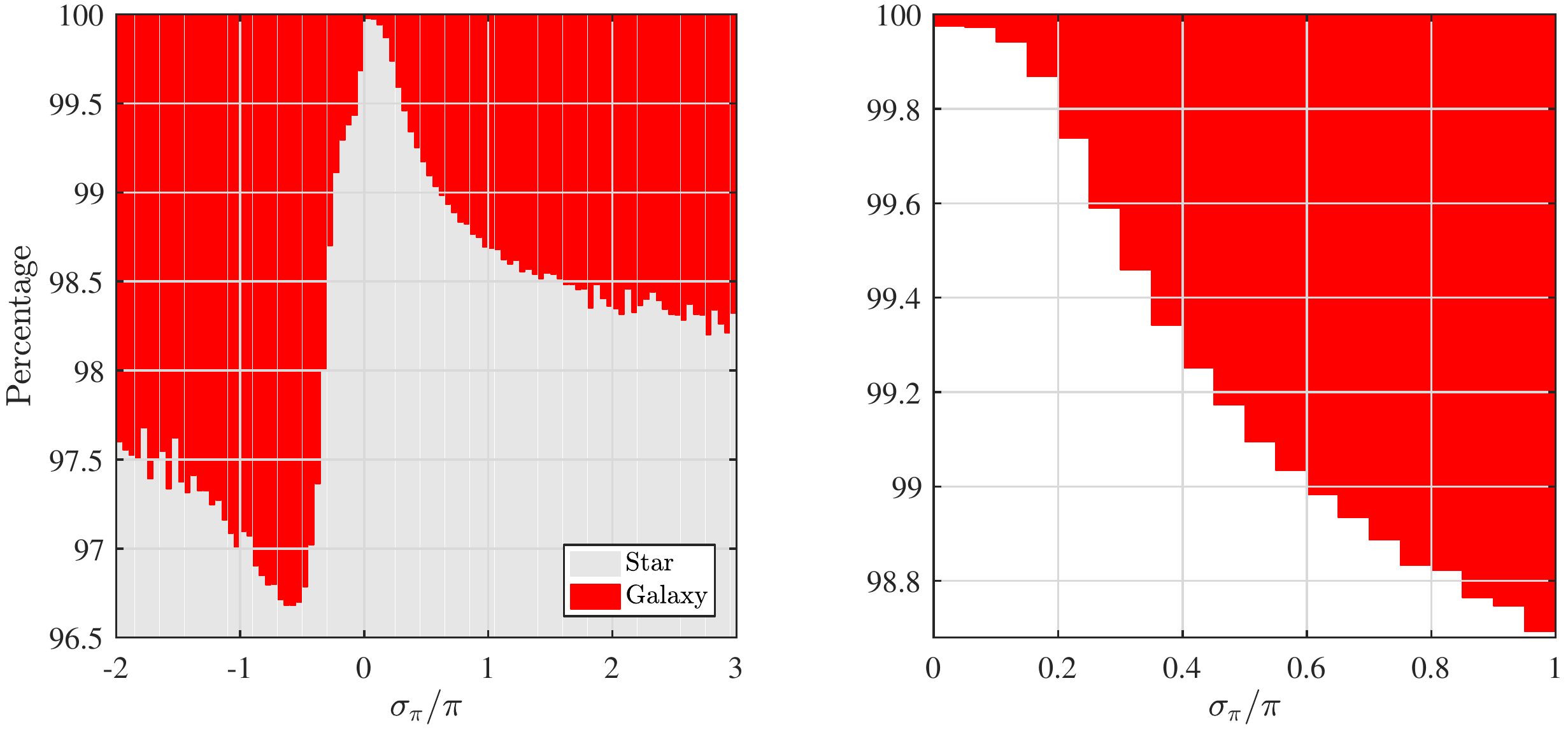}
  \caption{The stacked distributions of $\sigma_\pi/\pi$. Left panel: the distribution in the range between $-$2 and 3.
           Right panel: the distribution in the range between 0 and 1.  }
  \label{Contam2}
\end{figure}

%The Lutz$-$Kelker Effect (LKE) could cause the bias, when estimating distance directly from the trigonometric parallax \citep{Trumpler53,Lutz73}.
%Such bias depends on the relative parallax uncertainty.
The relative parallax uncertainty, $\sigma_{\pi}/\pi$, is an important parameter that can be used to constrain the bias caused by the Lutz$-$Kelker Effect
(LKE; \citealt{Trumpler53,Lutz73,Bai18b}).
We present the stacked distributions of $\sigma_{\pi}/\pi$ in Fig. \ref{Contam2}.
The sample of galaxies make up $\sim$2.5\% for the objects with parallaxes less than zero.
The percentages of the stars decrease sharply in the rang of $-$0.6 $< \sigma_\pi/\pi < $0.0, and reach the minimum 96.7\%.
Since there is no negative uncertainty, the negative $\sigma_\pi/\pi$ means negative parallax.

The percentages of the stars decline with the increase of the $\sigma_\pi/\pi$ for the objects with positive parallaxes.
The galaxies make up less than 1\% for the objects in the range of 0 $< \sigma_\pi/\pi <$ 0.6.
The sample are nearly all stars ($\sim$ 99.9\%) when 0 $< \sigma_\pi/\pi <$ 0.2.
In this range, the bias caused by the LKE also becomes insignificant \citep{Bai18b}.
Using the threshold of 0 $< \sigma_\pi/\pi <$ 0.2 could yield a very clean stellar sample,
%including  21,161,231 stars, 453 galaxies and 36 QSOs.
including 27,500,769 stars, 18,674 galaxies.

\section{Summary}\label{sum}
We apply the SL classifier to 85,613,922 objects in the $Gaia$ DR2,
based on the colors built from the PS1 and AllWISE.
%The criteria of the possibilities are set up with the help of the Simbad database, and there are over
%48 million objects left for the following studies.
The classification shows that about 98\% of the sample are stars, and 2\% are galaxies.
This result is cross-matched with Simbad database in order to estimate the
probability of the possibly wrong classifications, and the total accuracy is 91.9\%.
The Galactic plane is dominated by the stars, and the percentages become higher at high latitudes.
We find that about 2.5\% of the sample are galaxies for the objects with negative
parallaxes, and the threshold of 0 $< \sigma_\pi/\pi <$ 0.2 could yield a very clean stellar sample including
about 99.9\% stars.

\begin{acknowledgements}
This work was supported by the National Program on Key Research and Development
Project (Grant No. 2016YFA0400804) and
the National Natural Science Foundation of China (NSFC)
through grants NSFC-11603038/11333004/11425313/11403056.
Some of the data presented in this paper were obtained from the Mikulski Archive for
Space Telescopes (MAST). STScI is operated by the Association of Universities for Research
in Astronomy, Inc., under NASA contract NAS5-26555. Support for MAST for non-HST data is provided
by the NASA Office of Space Science via grant NNX09AF08G and by other grants and contracts.

This work has made use of data from the European Space Agency (ESA) mission
{\it Gaia} (\url{https://www.cosmos.esa.int/gaia}), processed by the {\it Gaia}
Data Processing and Analysis Consortium (DPAC,
\url{https://www.cosmos.esa.int/web/gaia/dpac/consortium}). Funding for the DPAC
has been provided by national institutions, in particular the institutions
participating in the {\it Gaia} Multilateral Agreement.
\end{acknowledgements}

\end{document}